\documentclass[conference]{IEEEtran}
\IEEEoverridecommandlockouts
\usepackage{cite}
\usepackage{amsmath,amssymb,amsfonts}
\usepackage{hyperref}
\usepackage{algorithmic}
\usepackage{graphicx}
\usepackage{textcomp}
\usepackage{xcolor}
\usepackage{amsmath}
\def\BibTeX{{\rm B\kern-.05em{\sc i\kern-.025em b}\kern-.08em
    T\kern-.1667em\lower.7ex\hbox{E}\kern-.125emX}}
\begin{document}

\title{
Rethinking IDE Customization for Enhanced HAX: A Hyperdimensional Perspective
}

\author{
    \IEEEauthorblockN{Roham Koohestani}
    \IEEEauthorblockA{
        Delft University of Technology\\
        rkoohestani@tudelft.nl
    }
    \and
    \IEEEauthorblockN{Maliheh Izadi}
    \IEEEauthorblockA{
        Delft University of Technology\\
        m.izadi@tudelft.nl
    }
}

\maketitle

\begin{abstract}
As Integrated Development Environments (IDEs) increasingly integrate Artificial Intelligence, Software Engineering faces both benefits like productivity gains and challenges like mismatched user preferences. We propose Hyper-Dimensional (HD) vector spaces to model Human-Computer Interaction, focusing on user actions, stylistic preferences, and project context. These contributions aim to inspire further research on applying HD computing in IDE design.
\end{abstract}


\section{Introduction}
\label{sec:intro}
Advancements in Artificial Intelligence (AI) have transformed Software Engineering (SE), with tools such as Cursor~\cite{cursor} and GitHub Spark~\cite{githubSpark} redefining development workflows. 
Despite gains in productivity and satisfaction~\cite{Kalliamvakou,jetbrains}, mismatches between developer preferences and AI-generated code persist~\cite{GitClear,sergeyuk2024ide}, leading to increased code churn.
Although fine-tuning can address these issues~\cite{technologies12110219}, it remains computationally expensive, leaving the customization of developer experiences largely unsolved. Existing research \cite{Parr2016towardsAUniversal} has explored the use of machine learning for automated code formatting, but these methods tend to incur substantial performance overhead. 

At the same time, an entire field of vector-symbolic AI has been gaining traction, with some methods leveraging it to store vast amounts of project context for model consumption~\cite{munkhdalai2024leavecontextbehindefficient}.
The field of vector-symbolic AI, and more broadly Hyper-Dimensional Computing (HDC), is not new. Its origins can be traced back to 1995, with Plate's development of holographic reduced representation~\cite{kanerva2009hyperdimensional,plate1995holographic}. 
Given the attributes of high-dimensional vector spaces, it is easy and efficient to model environments that store a large amount of context in a compact form. As there is a great amount of computational resource dependence for running state-of-the-art machine learning models, HDC can be seen as a less costly alternative for advanced behavior and preference modeling.

Next, we present the existing computational theory behind Hyper-Dimensional (HD) vector spaces and then explore their potential applications in 
customizable Human-AI Experience (HAX) designs.

\section{Hyperdimensional Computing}
\label{sec:hdc}
The term HDC was first coined by Pentti Kanerva~\cite{kanerva2009hyperdimensional} and builds on previous work by various others, such as Plate with Holographic Reduced Representation~\cite{plate1995holographic} and Gayler with Vector-Symbolic Architecture~\cite{gayler2004vector}. As the underlying theory is mostly the same, we will refer to the concept as HDC. We will specifically look at the Multiply Add Permute (MAP) framework by Gayler~\cite{gayler1998multiplicative};

\subsection{Foundations}
\label{sec:hdc:foudnations}
The MAP framework operates on hyper-dimensional vectors using three key operations: multiplication, addition, and permutation. These operations enable the composition, binding, and manipulation of high-dimensional representations.

\subsubsection{Variables}
\label{sec:hd:foundations:vars}
In this framework, variables are represented as randomly-sampled, high-dimensional, approximately orthogonal vectors. These vectors have the form $v \in \{-1, +1\}^D$. Although there are other variants of the framework with real- and integer-valued domains, we look at the bipolar variant. For instance, coding preferences like naming conventions or indentation styles could be defined as:
\[
\texttt{NameFormat} = V_1, \quad \texttt{Indentation} = V_2
\]
These variables are combined using the binding operation to represent more complex concepts holistically.

\subsubsection{Multiplication (Binding)}
\label{sec:hd:foundations:bind}
Binding involves combining two vectors to create a third vector that is dissimilar (orthogonal) to both. This ensures that information about the two input vectors is encoded in the resulting vector. Here, binding can be implemented as component-wise multiplication:
\[
\texttt{Bind}(A, B) = A \otimes B
\]
where \(A\) and \(B\) are sampled from $\{-1, +1\}^D$.
Note that the binding operator is also its own inverse.
% (the unbinding operator).

\subsubsection{Addition (Bundling)} 
\label{sec:hd:foundations:bundle}
Bundling aggregates multiple vectors into a single vector that represents their collective information. This is typically implemented as a component-wise sum followed by a normalization to remain in the same domain as the original vector:
\[
\texttt{Bundle}(A, B, C) = \text{Normalize}(A \oplus B \oplus C)
\]

\subsubsection{Permutation (Reordering)}
\label{sec:hd:foundations:permute}
Permutation denoted by \(P\) rearranges the elements of a vector to encode positional or structural information. For example, cyclically shifting vector components with respect to their position and subsequently bundling them can be used to denote a sequence:
\[
\texttt{Permute}(A) = P(A)
\]
% where \(P\) is a permutation operator.

\subsubsection{Similarity Analysis}
\label{sec:hd:similarity}
For evaluating the similarity of two vectors, the dot product/cosine similarity of the two vectors is calculated.

\section{HDC in IDEs}
\label{sec:hdc-in-ide}
We now shift focus to applying the theory of HDC to model two main aspects of any software project: user behavior/preferences and project context.

\subsection{Action Sequences - Next Action Prediction}
\label{sec:hdc-in-ide:nap}
To model sequences of user actions, we draw inspiration from the work of Mozannar et al.~\cite{mozannar2024readinglinesmodelinguser}, which focuses on modeling the states of developers. Consider an IDE that logs sequences of actions performed by a developer (e.g., opening files, typing code, running tests, etc.). Using HDC, we can represent these actions as high-dimensional vectors.

Each action (e.g., \texttt{OpenFile}, \texttt{RunTest}) can be represented as a vector sampled from the high-dimensional space. A sequence of $n$ actions is represented by binding and permuting these vectors to encode temporal order (see \autoref{sec:hd:foundations:permute}):
\[
\begin{aligned}
\texttt{Sequence} = & \, P^{n-1}(\texttt{Action1}) \otimes P^{n-2}(\texttt{Action2}) \\
                    & \, \otimes \ldots \otimes P^{0}(\texttt{ActionN})
\end{aligned}
\]
For example, if a user performs the actions \texttt{OpenFile}, \texttt{RunTest}, and \texttt{Commit}, the sequence can be encoded as:
\[
\begin{aligned}
\texttt{Sequence} = & \, P^2(\texttt{OpenFile}) \otimes P^1(\texttt{RunTest}) \\
                    & \, \otimes P^0(\texttt{Commit})
\end{aligned}
\]
For a sequence of $M$ actions where $m \geq n$ we can encode a user's behavior $UB$ as
\[
\bigoplus_{i=0}^{m-n} \texttt{encode}((\texttt{Action}_i, \ldots, \texttt{Action}_{i+n}))
\]

To predict the next action after observing a sequence of $n-1$ actions, we use the properties of HD vector spaces. As the binding operation is distributive over bundling, we can attempt to extract the next action by applying $UB \otimes P(\texttt{encode}(\texttt{Action}_1, \ldots, \texttt{Action}_{n-1}))$. As all other dissimilar vectors result in negligible noise, the remaining vector $\texttt{PredAcc}$ will be highly similar to the vector of the next action of the user. This action $\texttt{ActionX}$ can therefore be found as 

\[
\arg\max_{\texttt{ActionX}} \text{Similarity}(\texttt{PredAcc}, \texttt{ActionX})
\]
This allows the IDE to predict the next most likely action, enabling the optimization of the user's experience.

\subsection{Stylistic Preferences - Style-Matched Generation}
\label{sec:hdc-in-ide:smg}
Developers often have personal stylistic preferences when writing code. HDC can model and enforce these preferences for tasks like code completion~\cite{izadi2024language,izadi2022codefill} or auto-formatting~\cite{prabhu2017dynamic}.

Using the approach inspired by Kanerva's ``Dollar of Mexico'' analogy~\cite{kanerva2010we}, we encode stylistic preferences for different languages or individual developers. For instance:
\[
\begin{aligned}
\texttt{STYLE} = & \, (\texttt{NameFormat} \otimes \texttt{CamelCase}) \\
                 & \, \oplus (\texttt{Indentation} \otimes \texttt{Spaces4})
\end{aligned}
\]
\[
\begin{aligned}
\texttt{MODEL\_STYLE} = & \, (\texttt{NameFormat} \otimes \texttt{SnakeCase}) \\
                        & \, \oplus (\texttt{Indentation} \otimes \texttt{Tabs})
\end{aligned}
\]

To adapt the generated code to the developer's style, a mapping vector is created:
\[
\texttt{MAPPING} = \texttt{MODEL\_STYLE} \otimes \texttt{STYLE}
\]
If a LLM generates code with \texttt{NameFormat = SnakeCase}, the mapping ensures it is translated to \texttt{CamelCase}:
\[
\texttt{CamelCase} \approx \texttt{SnakeCase} \otimes \texttt{MAPPING}
\]

This enables the IDE to generate dynamically style-matched code and maintain consistent project styling.

\subsection{Representing Project Context}
\label{sec:hdc-in-ide:rpc}
HDC also provides a robust framework for modeling the context of a software project and encompasses aspects such as programming languages,  Application Programming Interface (APIs), design patterns, and usage scenarios.

For example, the project's context can be encoded as:
\[
\begin{aligned}
\texttt{CONTEXT} = & \, (\texttt{LANG} \otimes \texttt{Python}) \\
                   & \, \oplus (\texttt{API} \otimes \texttt{TensorFlow}) \\
                   & \, \oplus (\texttt{Pattern} \otimes \texttt{Observer})
\end{aligned}
\]
This holistic vector representation allows the IDE to adapt suggestions and auto-completions to the specific context of the project. For instance, when working on a Python project with TensorFlow, the IDE can prioritize TensorFlow-related completions or suggest design patterns suitable for Python.

Furthermore, transitions between contexts, such as switching from hobby projects to work-related projects, can be modeled using mappings represented as:
\[
\texttt{WORK\_CONTEXT} * \texttt{HOBBY\_CONTEXT}
\]

\section{Future Direction}
\label{sec:future-direction}
While we have outlined several ways HDC can help model user behaviors to enhance their experience, the challenge lies in applying these methods effectively in real-world scenarios.
Future research should explore methods to incorporate these representations more effectively and ensure that they influence the models' generations.
There exists research looking at interventions at the decoding stage \cite{dai2024mpcodermultiuserpersonalizedcode} to improve the coding style adherence of LLMs. 
Additionally, IDE developers could explore adopting a mapping approach similar to the model-to-user mapping discussed in the previous section. This would allow them to align the style of code generated by the LLM with the user's preferred coding style.

\section{Conclusion}
\label{sec:conclusion}
In this paper, we look at applying the HDC theory to modeling user behavior and preferences through Hyper-dimensional vectors. We present three use cases in which user actions, stylistic preferences, and project setup can be represented using HDC. We encourage the IDE research and development community to engage in research that attempts to include such representations inside the IDE to improve the experience of users in the IDE through efficient approaches.

\bibliographystyle{IEEEtran}
\bibliography{main}

\end{document}